\documentclass[aps,prstab,onecolumn,groupedaddress]{revtex4-1}

\usepackage{graphicx}
\usepackage{subfigure}

\usepackage{natbib}
\begin{document}


\title{Transfer Matrix for combined RF and Solenoid Fields}


\author{Colwyn Gulliford, Ivan Bazarov}
\affiliation{CLASSE, Cornell University}


\date{\today}

\begin{abstract}
We present a new method for computing the transverse transfer matrix through superimposed axisymmetric RF and solenoid field maps.  The algorithm constructs the transfer matrix directly from one dimensional RF and solenoid field maps without computing numerical derivatives or eigenfunction expansions of the field map data.  In addition, this method accurately describes the dynamics of low energy particles starting from a solenoid immersed cathode, allowing the method to be used to simulate transport through both RF and electrostatic guns.  Comparison of particle tracking with the transfer matrix and direct integration of the equations of motion through several field set-ups shows excellent agreement between the two methods.

\end{abstract}

\pacs{PACS numbers?}


\maketitle

\section{Introduction\label{intro}}
				
Linear transfer matrices continue to serve as important simulation tools in the design, commissioning, and operation of modern accelerators.  Common examples of their use include computing linear centroid motion, beam based alignment of optical elements, and orbit feedback.   In addition, they also feature prominently in the theory of round-to-flat beam transforms in RF and DC guns \cite{debby,kkim}, and are used for emittance measurements for beams without space charge \cite{ivan1}.  Because of this utility, analytic expressions for the transfer matrix through many beam line elements, such as magnets with constant fields, are well known and are widely used.  In contrast to these simple elements, the fields of many beam line elements in modern accelerators have no analytic form and may overlap each other.  For example, in high brightness electron sources, solenoid fields used for emittance compensation may overlap the accelerating fields at or near the cathode.   To properly describe the dynamics in these machines, the transfer matrix through superimposed RF and solenoid fields must be constructed.  In general, to model these elements one must use numerically computed electromagnetic field maps.  Unfortunately, no closed form solutions for the transfer matrix through such elements exist.

Nonetheless, a significant amount of work has gone into developing both semi-analytic and numerical techniques to compute these matrices.  In general, these techniques require some form of manipulation of the field map data and often have a limited range of validity.  For example, the widely used RF transfer matrix given by Rozensweig and Serafini \cite{rosy} requires a Floquet expansion of the on-axis RF field map, and is only valid for ultra-relativistic particles.  Other methods expand the field map data in terms of the the general solution to the homogeneous Maxwell equations in cylindrical coordinates \cite{chao,abell}.  From this expansion, the vector potential can be computed, allowing the Hamiltonian for the overlapping solenoid and RF fields to be constructed.  When combined with differential and lie algebra techniques this method is quite powerful, and can be used to generate arbitrary order maps.  Despite this, it requires significant overhead in setting up and may not be suitable for online modeling purposes.  A simpler solution has been put forth in \cite{yu}.  In this approach, the particle energy and RF fields are assumed constant over a small time step.  The second order linear transverse equations of motion can then be solved `exactly', and the transfer matrix for the time step constructed.  While this method works well for relativistic particles, it does have  several draw backs.  First, the assumption that the energy is constant requires extremely small time steps for very low energy particles like those emitted from a photocathode.  Secondly, the determinant of the resulting transfer matrix is only correct to first order in the step size.  

Depending on the particular application, one or more of these methods may be appropriate to use.  However, the need for a simple algorithm that computes the transfer matrix through superimposed RF and solenoids fields accurately for low particle energies still exists.  In order to be useful, such a matrix should satisfy the following three requirements: (i) using reasonable step sizes, the matrix should be able to describe the low energy dynamics found in RF or DC guns, (ii) the matrix should have the correct determinant regardless of the step size, (iii) the algorithm for constructing the matrix should be easy to implement.  Based on a new solution to the equations of motion, we derive a transfer matrix with all three of these qualities.  

The layout of this work is as follows.  First, the longitudinal equations of motion are solved for a single small step in the independent variable.  Then, the transfer matrix over the same step is derived for electrostatic and solenoid fields.  Building on this result, the matrix for combined RF and solenoid fields is computed.  This matrix is then tested with tracking through a DC gun, a superconducting RF cavity and a RF gun with a solenoid immersed cathode.  Excellent agreement between the transfer matrix and direct integration of the equations of motion is demonstrated in all three cases.




\section{Field Expansions and the Equations of Motion}

In general, both standing and traveling wave RF fields can be written in the complex form
${\mathbf{E}} ={\mathbf{\tilde\mathcal{E}}}(x,y,z)e^{i\omega t}$ and ${\mathbf{B}} = \tilde{\mathbf{\mathcal{B}}}(x,y,z) e^{i\omega t}$.  
In these and all subsequent expression, tildes are used to denote phasor quantities.  The coordinate system for the fields is set up so that the $z$-axis points along the length of the beamline.  The functions $\tilde{\mathcal{E}}$ and $\tilde{\mathcal{B}}$ represent the field maps generated by RF cavity field solvers, and include the initial phase offset of the cavity.  For notational simplicity, the phase factor $e^{i\omega t}$ and the real symbol $\mathrm{Re}[...]$ are suppressed in all but the final results for the transfer matrix.  To derive the linear equations of motion, the fields are expanded to first order in the transverse offsets $x$ and $y$.  The linearized RF fields take the form
\begin{eqnarray}
\tilde{\mathcal{E}}(x,y,z) =  -\frac{x}{2}\left(\frac{d\tilde{\mathcal{E}}_z}{dz}\right)\hat{\mathbf{x}}
 -\frac{y}{2}\left(\frac{d\tilde{\mathcal{E}}_z}{dz}\right)\hat{\mathbf{y}}
+ \tilde{\mathcal{E}}_z(r=0)\hat{\mathbf{z}},
\nonumber
\\
\tilde{\mathcal{B}}(x,y,z) = -\frac{y}{2}\left(\frac{i\omega}{c^2}\right)\tilde{\mathcal{E}}_z\hat{\mathbf{x}}
+ \frac{x}{2}\left(\frac{i\omega}{c^2}\right)\tilde{\mathcal{E}}_z\hat{\mathbf{y}}.
\end{eqnarray}
Similarly, the expansion of the solenoid field gives
\begin{eqnarray}
\mathbf{B}_{\mathrm{sol}} = 
-\frac{x}{2}\left(\frac{dB_z}{dz}\right)\hat{\mathbf{x}}
-\frac{y}{2}\left(\frac{dB_z}{dz}\right)\hat{\mathbf{y}} + B_z(r=0)\hat{\mathbf{z}}.
\end{eqnarray}
Note that the solenoid field is distinguished from the RF magnetic field by the lack of a tilde.
The total physical fields are given by $\mathbf{E}_{\mathrm{tot}} = \mathrm{Re}[\tilde{\mathcal{E}} e^{i\omega t}]$ and
$\mathbf{B}_{\mathrm{tot}} = \mathrm{Re}[\tilde{\mathcal{B}} e^{i\omega t}]  + \mathbf{B}_{\mathrm{sol}}$.

For the single particle equations of motion, the longitudinal coordinate $z$ is used as the independent variable.  Derivatives with respect to $z$ are denoted with a prime: $f^{\prime} \equiv df/dz$.  The equations of motion for the longitudinal phase space variables $t$ and $\gamma$ are
\begin{eqnarray}
t^{\prime} = \frac{1}{\beta(z)c},\hspace{1cm}
\tilde{\gamma}^{\prime}(z) = \frac{e\tilde{\mathcal{E}}_z}{mc^2}
= \tilde{\mathcal{E}}_z/\mathcal{E}_e, \hspace{1cm} \gamma^{\prime}  = \mathrm{Re}[\tilde\gamma^{\prime}e^{i\omega t}].
\label{LongEOM}
\end{eqnarray}
Here the constant $\mathcal{E}_e = mc^2/e$ gives the (signed) rest energy of the electron in [eV].  The normalized energy is given by $\gamma = \mathrm{Re}[\tilde{\gamma}e^{i\omega t}]$.  
The transverse equations of motion can be written as \cite{riser}:
\begin{eqnarray}
x^{\prime\prime}+\frac{p^{\prime}}{p}x^{\prime}
+\left(\frac{\gamma\tilde{\mathcal{E}}^{\prime}_z}{2\mathcal{E}_ep^2}
+\frac{i\omega\tilde{\gamma}^{\prime}}{2cp}\right)x
+2\Delta\theta_L^{\prime}y^{\prime}
-\left(\frac{cB_z^{\prime}}{2\mathcal{E}_ep}\right)y &=&  0, 
\label{GenEOM1}
\\
y^{\prime\prime}+\frac{p^{\prime}}{p}y^{\prime}
+\left(\frac{\gamma\tilde{\mathcal{E}}^{\prime}_z}{2\mathcal{E}_ep^2}
+\frac{i\omega\tilde{\gamma}^{\prime}}{2cp}\right)y
-2\Delta\theta_L^{\prime}x^{\prime}
+\left(\frac{cB_z^{\prime}}{2\mathcal{E}_ep}\right)x &=&  0. 
\label{GenEOM2}
\end{eqnarray}
In this expression, $p=\beta\gamma$ is the normalized reference particle momentum, and $\Delta\theta_L$ is the Larmor angle defined by
\begin{eqnarray}
\Delta\theta_L(z) = 
- \int_{z_i}^z\frac{cB_z}{2\mathcal{E}_ep}dz,\hspace{1cm}\Delta\theta_L^{\prime} = -\frac{cB_z}{2\mathcal{E}_ep}.
\end{eqnarray}
Note the use of a negative sign in front of the integral.  With this definition, a positive solenoid field creates a positive change in the Larmor angle for electrons.  Decoupling these equations requires rotating to the Larmor frame.  By defining the variable ${\eta} = x+iy$, the transformation to Larmor coordinates takes the form \cite{riser}:
\begin{eqnarray}
{\eta}_L(z) = x_L+iy_L = {\eta} e^{-i\Delta\theta_L(z)}.
\label{larmorTrans}
\end{eqnarray}
The equivalent transformation for the transverse phase space vector 
$\mathbf{u} = \left( 
\begin{array}{cccc}
x  & x^{\prime} & y & y^{\prime}
\end{array} 
\right)^{\mathrm{T}}$ is defined by
\begin{eqnarray}
\left( 
\begin{array}{c}
x_L  \\
x^{\prime}_L \\
y_L \\
y^{\prime}_L
\end{array} 
\right)
=
\left( 
\begin{array}{cccc}
C & 0 & S & 0 \\
-\Delta\theta_L^{\prime}S & C & \Delta\theta_L^{\prime}C & S\\
-S & 0 & C & 0 \\
-\Delta\theta_L^{\prime}C & -S & -\Delta\theta_L^{\prime}S  & C
\end{array} 
\right)
\left( 
\begin{array}{c}
x  \\
x^{\prime} \\
y \\
y^{\prime}
\end{array} 
\right).
\end{eqnarray}
In this and following expressions, $C \equiv \cos\Delta\theta_L$, and $S\equiv\sin\Delta\theta_L$.  The matrix in this equation is denoted by $L = L(\Delta\theta_L,\Delta\theta_L^{\prime})$.  It is easy to show that the determinant of this matrix is equal to unity.  Another important characteristic of this matrix is found by taking the limit as $z\rightarrow z_i$.  In this limit, $\Delta\theta_L\rightarrow 0$, and the matrix reduces to
\begin{eqnarray}
L(0,\Delta\theta_L^{\prime}(z_i))=
\left( 
\begin{array}{cccc}
1 & 0 &0 & 0 \\
0 &1 & \Delta\theta_L^{\prime}(z_i) & 0 \\
0 & 0 & 1 & 0 \\
-\Delta\theta_L^{\prime}(z_i) & 0  & 0 & 1
\end{array} 
\right).
\end{eqnarray}
The two remaining focusing terms describe the effect of starting a particle with $B_z\neq 0$, the so called `immersed cathode' condition.

Substituting the transformation in  Eq.~(\ref{larmorTrans}) into Eqs.~(\ref{GenEOM1}) and (\ref{GenEOM2}) gives the transverse equation of motion for the Larmor coordinates:
\begin{eqnarray}
\eta_L^{\prime\prime}+\frac{p^{\prime}}{p}\eta_L^{\prime}+\left(\frac{\gamma\tilde{\mathcal{E}}_z^{\prime}}{2\mathcal{E}_ep^2}
+\frac{i\omega\tilde{\gamma}^{\prime}}{2cp} + (\Delta\theta_L^{\prime})^2\right)\eta_L = 0.
\label{LarmorEOM}
\end{eqnarray}
In this expression $\eta_L$ stands for either $x_L$ and $y_L$.  Because the transformation between the laboratory and Larmor coordinates is known, all that remains is to solve the above differential equation. With this solution, the transfer matrix for the lab phase space variables follows directly:  $M(z_i\rightarrow z_f) = L^{-1}(z_f)M_L(z_i\rightarrow z_f)L(z_i)$.  Consequently, the majority of this work is spent finding an exact solution to Eq.~(\ref{LarmorEOM}) within the approximation that the fields are constant over a small step in $z$.


\section{Derivation of the Transfer Matrix}

As stated before, general solutions to the differential equations in Eqs.~(\ref{LongEOM}-\ref{GenEOM2}), and (\ref{LarmorEOM}) do not exist for arbitrary cavity and solenoid field maps.  As a result, any method for computing the transfer matrix analytically requires some form of approximation to these equations.  The approach taken in this work is to find an exact solution to the equations of motion for a step in $z$ and $t$ that is small enough so that the fields profiles $\tilde\mathcal{E}_z(z)$ and $\tilde\mathcal{B}_z(z)$, as well as the RF phase, don't change appreciably.  The solution is exact in the sense that the particle energy changes correctly over the course of the step.  The change in the field map profile as well as the RF phase are then included with the use of edge focusing matrices.  Slicing the field maps and consecutively multiplying the matrices for each step gives the total transfer matrix:
\begin{eqnarray}
M(z_i\rightarrow z_f) \approx \prod_k \Delta M(z_{k}\rightarrow z_{k}+\Delta z).
\end{eqnarray}

The first step in constructing the transfer matrix for one step is to solve the longitudinal equations of motion in Eq.~(\ref{LongEOM}).  For a constant electric field, $\gamma^{\prime} $ is constant, and the normalized energy, momentum, and velocity are given by
\begin{eqnarray}
\gamma(z) = \gamma_i + \gamma^{\prime}(z-z_i),
\hspace{1cm} p(z) = \sqrt{(\gamma_i+\gamma^{\prime}(z-z_i))^2-1},
\hspace{1cm} \beta(z) = p(z)/\gamma(z).
\label{gamma}
\end{eqnarray}
Using these expressions, the derivatives of $t(z)$ and $\Delta\theta_L$(z), in Eqs.~(\ref{LongEOM}) and (\ref{LarmorEOM}) can be directly integrated:
\begin{eqnarray}
t(z) &=& \frac{1}{c\gamma^{\prime}}[p(z)-p_i],
\label{Teqn}
\\
\Delta\theta_L(z) &=& \left(\frac{b}{\gamma^{\prime}}\right)\ln\left(\frac{p(z)+\gamma(z)}{p_i+\gamma_i}\right).
\label{Leqn}
\end{eqnarray}
In the last line, the constant $b$ is defined as $b = p\cdot\Delta\theta_L^{\prime}=-eB_z/2mc$.   In addition to defining the transformation between the lab and Larmor coordinates, the function $\Delta\theta_L$ plays an important role in the derivation of the transfer matrix for both the electrostatic and RF field cases.

\subsection{Overlapping Electrostatic and Solenoid Fields}

The derivation of the transfer matrix for superimposed RF and solenoid fields is based in part on the method used to derive the transfer matrix for static fields \cite{ivan1}.  In this section, a detailed derivation of the static field result is given.  The same techniques are then modified and used to derive the RF matrix in the following section.  For static fields, the equation of motion is found by taking $\omega\rightarrow 0$ in Eq.~(\ref{LarmorEOM}):
\begin{eqnarray}
\eta_L^{\prime\prime}+\frac{p^{\prime}}{p}\eta_L^{\prime}+\left(\frac{\gamma\mathcal{E}_z^{\prime}}{2\mathcal{E}_ep^2}
+ (\Delta\theta_L^{\prime})^2\right)\eta_L = 0.
\end{eqnarray}
Note that $p^{\prime}\propto \mathcal{E}_z$ and $\Delta\theta_L^{\prime}\propto B_z$ in this expression, implying $\eta_L$ depends on both the accelerating field and its gradient, as well as the solenoid field.  The transfer matrix from $z_i$ to $z_f = z_i+\Delta z$ is derived in a three step process.  Over the interval $[z_i,z_f]$, the electric and magnetic fields are approximated as rectangular step functions.  Formally the fields are written as:
\begin{eqnarray}
\mathcal{E}_z(z) &\cong& \mathcal{E}_z(z_i)\left\{\theta(z-z_i)-\theta(z-z_f)\right\},
\hspace{1cm}
\mathcal{E}_z^{\prime} \cong \mathcal{E}_z(z_i)\left\{\delta(z-z_i) - \delta(z-z_f)\right\},
\nonumber
\\
B_z(z) &\cong& B_z(z_i)\left\{\theta(z-z_i)-\theta(z-z_f)\right\},
\hspace{1cm}
\hspace{-0.15cm}B_z^{\prime} \cong B_z(z_i)\left\{\delta(z-z_i) - \delta(z-z_f)\right\},
\end{eqnarray}
where $\theta(z)$ is the Heaviside step function and $\delta(z)$ is the Dirac delta function.  The transfer matrix is then found by solving the transverse equation of motion piecewise from $z_i$ and $z_f$.

First, the equations of motion are integrated across the rising edge of the electric field at $z_i$.  Because the rising edge is approximated as an instantaneous step, the particle's position does not changed: $\eta_L(z_i^+)= \eta_L(z_i^-)=\eta_L(z_i)$.  Integrating the equation of motion gives the kick delivered to the particle's trajectory:
\begin{eqnarray}
\Delta\eta_L^{\prime} &=& 
-\int_{z_i-\epsilon}^{z_i+\epsilon}\left\{\frac{p^{\prime}}{p}\eta_L^{\prime}+\left(\frac{\gamma\mathcal{E}_z^{\prime}}{2\mathcal{E}_ep^2}
+ (\Delta\theta_L^{\prime})^2\right)\eta_L\right\}dz
 = -\int_{z_i-\epsilon}^{z_i+\epsilon}
\frac{\mathcal{E}_z(z_i)}{2\gamma\beta^2\mathcal{E}_e}\eta_L(z)\delta(z-z_i)dz=-\frac{\gamma^{\prime}}{2\gamma_i\beta_i^2}\eta_L(z_i),
\nonumber
 \end{eqnarray}
The corresponding transfer matrix for the rising edge, $R_E$, takes the form
\begin{eqnarray}
R_{E}(\gamma,\gamma^{\prime}) = 
\left( 
\begin{array}{cc}
1 & 0 \\
-\frac{\gamma^{\prime}}{2\gamma\beta^2} & 1
\end{array} 
\right).
\end{eqnarray}

Next, the equation of motion is solved across the interval $(z_i,z_f)$ where both the electric and magnetic field are approximately constant.   In this region, the equation of motion reduces to
\begin{eqnarray}
\eta_L^{\prime\prime}+\frac{\gamma^{\prime}}{\gamma\beta^2}\eta_L^{\prime}+(\Delta\theta_L^{\prime})^2\eta_L=0,
\label{eomDC}
\end{eqnarray}
where $\Delta\theta_L^{\prime}(z) \propto 1/p(z)$.  With the electric field held constant, $\gamma$, $p$, $\beta$, and $\Delta\theta_L$ are given by Eqs.~(\ref{gamma}) and (\ref{Leqn}).  With these functions, the differential equation can be solved by assuming $\eta_L=\eta_L(\Delta\theta_L)$.  Plugging this into Eq.~(\ref{eomDC}) gives
\begin{eqnarray}
(\Delta\theta^{\prime}_L)^2\left[\frac{d^2\eta_L}{d\theta_L^2}+\eta_L\right] 
+ \frac{d\eta_L}{d\theta_L}\left[\Delta\theta_L^{\prime\prime}+\frac{\gamma^{\prime}}{\gamma\beta^2}\Delta\theta_L^{\prime}\right] = 0.
\end{eqnarray}
Using Eq.~(\ref{Leqn}), is is possible to show $\Delta\theta^{\prime\prime} = -\gamma^{\prime}\Delta\theta^{\prime}/\gamma\beta^2$, canceling the second term the above equation.  Assuming $\Delta\theta^{\prime}\neq0$, the first term in this expression must also vanish. It follows that: $\eta_L = A\cos\Delta\theta_L + B\sin\Delta\theta_L$.  
Completing the initial value problem for this solution determines the transfer matrix for the step from $z_i$ to $z_f$:
\begin{eqnarray}
M_{i\rightarrow f} = \left( 
\begin{array}{cc}
C &\frac{p_i}{b}S \\
-\frac{b}{p_f}S & \frac{p_i}{p_f}C
\end{array} 
\right).
\end{eqnarray}

The last step in constructing the full matrix for the interval $[z_i,z_f]$ is to evaluate the transfer matrix for the falling edge of the accelerating field.  The result is essentially the same as before, except now the derivative of the electric field has the opposite sign.  This allows the transfer matrix for the falling edge to be written as
\begin{eqnarray}
R_E^{-1}(\gamma,\gamma^{\prime}) = 
\left( 
\begin{array}{cc}
1 & 0 \\
\frac{\gamma^{\prime}}{2\gamma\beta^2} & 1
\end{array} 
\right).
\end{eqnarray}
Combining the three matrices for each region gives the full transfer matrix for the step $\Delta z$:
\begin{eqnarray}
\Delta M^{\mathrm{dc}}_{x,x^{\prime}} = 
R_E^{-1}(\gamma_f,\gamma^{\prime})M_{i\rightarrow f}R_E(\gamma_i,\gamma^{\prime})=
\left( 
\begin{array}{cc}
1 & 0 \\
\frac{\gamma^{\prime}}{2\gamma_f\beta_f^2} & 1
\end{array} 
\right)
 \left( 
\begin{array}{cc}
C &\frac{p_i}{b}S\\
-\frac{b}{p_f}S & \frac{p_i}{p_f}C
\end{array} 
\right)
\left( 
\begin{array}{cc}
1 & 0 \\
-\frac{\gamma^{\prime}}{2\gamma_i\beta_i^2} & 1
\end{array} 
\label{DCmatrix}
\right).
\end{eqnarray}
One important thing to note about the matrix in Eq.~(\ref{DCmatrix}) is that it has the correct determinant for the phase space variables chosen: $\mathrm{det}(\Delta M^{\mathrm{dc}}_{x,x^{\prime}}) = p_i/p_f$.  The transfer matrix for the canonical phase space variables $x_L$ and $p_{x,L}$ can be found by applying the transformation:
\begin{eqnarray}
\Delta M^{\mathrm{dc}}_{x,p_x} = \left( 
\begin{array}{cc}
1 & 0 \\
0 & p_f
\end{array} 
\right)\Delta M^{\mathrm{dc}}_{x,x^{\prime}} 
\left( 
\begin{array}{cc}
1 & 0 \\
0 & 1/p_i
\end{array} 
\right),
\end{eqnarray}
It follows from this expression that the matrix $\Delta M^{\mathrm{dc}}_{x,p_x}$ satisfies the symplectic condition $\mathrm{det}(\Delta M^{\mathrm{dc}}_{x,p_x}) = 1$.  In addition to this, the transfer matrix also has the convenient feature that the derivative of the accelerating field never has to be calculated, bypassing the need to compute derivatives numerically.  It is important to note that when starting a particle from a non-zero electric field, the first rising edge matrix should not be included .  Doing so amounts to the particle seeing the field rise from zero to the actual value of the field at the cathode and is not physical.  Similarly, the falling edge matrix at the end of tracking should not be included when the particle is found in a non-zero electric field. 

\subsection{Overlapping RF and Solenoid Fields}

With the results for the electrostatic field worked out, it is now possible to construct the transfer matrix for RF fields.  The approximation used here is similar to that used in the electrostatic case: the field profiles $\tilde{\mathcal{E}}_z$ and $B_z$, as well as the RF phase, are assumed constant over the step $\Delta z$.  This implies $\tilde\gamma(z,t)$ is  also constant, allowing the solutions to the longitudinal equations of motion in Eq.~(\ref{gamma}) to be used.  The effect of adding the time dependence to the edge matrices is minimal;  the rising edge matrix remains the same.  For the falling edge matrix, the change in the RF phase is included in the electric field: $\tilde\gamma_f^{\prime} = \tilde\gamma^{\prime}e^{i\omega\Delta t}$.  Note that this reduces to the electrostatic case when $\omega\rightarrow 0$.  The equation of motion over the interval $(z_i,z_f)$ reduces to
\begin{eqnarray}
\eta_L^{\prime\prime}+\frac{\tilde\gamma^{\prime}}{\gamma\beta^2}\eta_L^{\prime}
+\left(\frac{i\omega\tilde{\gamma}^{\prime}}{2cp} +(\Delta\theta_L^{\prime})^2\right)\eta_L = 0.
\end{eqnarray}
Unfortunately, the extra focusing from the RF magnetic field scales as $p^{-1}$, and the form of the solution used in the electrostatic case no longer solves the equation of motion.  To our knowledge, there is no solution to this equation in its current form.

In order to solve this equation, it must be transformed in such a way that the RF magnetic focusing term is removed.  To do so requires switching the independent variable from longitudinal position to time.  The resulting transformation of the phase space variables takes the form
\begin{eqnarray}
\left( 
\begin{array}{c}
\eta \\
\dot\eta 
\end{array} 
\right)
=
T
\left( 
\begin{array}{c}
\eta \\
\eta^{\prime} 
\end{array} 
\right),
\hspace{1cm}
T(\beta) = 
 \left( 
\begin{array}{cc}
1 & 0 \\
0 & c\beta
\end{array} 
\right).
\end{eqnarray}
The transverse equation of motion with time as the independent variable is given by
\begin{eqnarray}
\ddot\eta + \frac{\dot{\tilde\gamma}}{\gamma}\dot\eta + \left[ \frac{c^2\tilde{\mathcal{E}}_z^{\prime}}{2\gamma \mathcal{E}_e}
+ c\beta\frac{i\omega\tilde\gamma^{\prime}}{2\gamma}  + (\Delta\dot{\theta}_L)^2 \right]\eta = 0. 
\end{eqnarray}
The second transformation is to reduce this equation by introducing the variable $\hat{\eta}_L = \sqrt{\gamma}\eta_L$.  Once again, this transformation can be written in matrix form:
\begin{eqnarray}
\left( 
\begin{array}{c}
\hat{\eta}_L \\
\dot{\hat{\eta}}_L
\end{array} 
\right)
=
\Lambda
\left( 
\begin{array}{c}
\eta_L \\
\dot\eta_L
\end{array} 
\right),
\hspace{1cm}
\Lambda(\gamma,\dot{\tilde\gamma}) = 
\sqrt{\gamma} \left( 
\begin{array}{cc}
1 & 0 \\
\frac{\dot{\tilde\gamma}}{2\gamma} & 1
\end{array} 
\right).
\end{eqnarray}
The equation of motion for the reduced variables takes the form of Hill's equation:
\begin{eqnarray}
\ddot{\hat\eta} + \left[\dot p^2\frac{(p^2-2)}{4(p^2+1)^2}
+\frac{c^2\tilde{\mathcal{E}}_z^{\prime}}{2\gamma^3\mathcal{E}_0}+(\Delta\dot{\theta}_L)^2\right]\hat\eta = 0.
\label{redEOM}
\end{eqnarray}
The important thing to notice about this expression is that the problematic RF focusing term has been hidden in the variable transformation.  All that remains now is to solve this equation in the region where the solenoid and accelerating fields are constant.  To do so requires knowing the functions $p(t)$ and $\gamma(t)$.  The momentum is easily found by rearranging Eq.~(\ref{Teqn}): $p(t) = p_i + c\gamma^{\prime}(t-t_i)$.  The normalized energy is then given by  $\gamma(t) = \sqrt{p^2(t)+1}$.  Inserting these expressions into the equations of motion in the constant field region yields:
\begin{eqnarray}
\ddot{\hat\eta} + \left[(c\tilde\gamma^{\prime})^2\frac{(p^2-2)}{4(p^2+1)^2}
+\frac{(bc)^2}{p^2+1}\right]\hat\eta = 0.
\end{eqnarray}
This equation can be solved with the function $\hat{n}_L = \sqrt{\gamma}(A\cos\Delta\theta_L + B\sin\Delta\theta_L)$.  The transfer matrix is found by completing the initial value problem for this solution.  The resulting matrix can be written in the compact form
\begin{eqnarray}
\hat{M}_{i\rightarrow f}^{\mathrm{rf}} = \Lambda(\gamma_f,\dot{\tilde\gamma}_f)
T(\beta_f)M_{i\rightarrow f}^{\mathrm{dc}}T^{-1}(\beta_i)\Lambda^{-1}(\gamma_i,\dot{\tilde\gamma}).
\end{eqnarray}
In this expression $\dot{\tilde\gamma}_f= c\beta_f\tilde\gamma^{\prime}$.  This matrix correctly describes the evolution of the reduced variables.  In order to get the transfer matrix for the usual phase space variables it must be transformed back:
\begin{eqnarray}
M_{i\rightarrow f}^{\mathrm{rf}} = T^{-1}(\beta_f)\Lambda^{-1}(\gamma_f,\dot{\tilde\gamma}_fe^{i\omega\Delta t})\Lambda(\gamma_f,\dot{\tilde\gamma}_f)
T(\beta_f)M_{i\rightarrow f}^{\mathrm{dc}}.
\end{eqnarray}
It is clear from this expression that the effects of the RF magnetic focusing must be contained in the matrices left multiplying $M_{i\rightarrow f}^{\mathrm{dc}}$.  Combining these matrices together gives
\begin{eqnarray}
R_{\mathrm{rf}}=T^{-1}(\beta)\Lambda^{-1}(\gamma,\dot{\tilde\gamma} e^{i\omega\Delta t})\Lambda(\gamma,\dot{\tilde\gamma})
T(\beta) =  \left( 
\begin{array}{cc}
1 & 0 \\
\frac{\tilde\gamma^{\prime}}{2\gamma}(1-e^{i\omega\Delta t})& 1
\end{array} 
\right)
\end{eqnarray} 
In the limit that $\Delta t = t-t_i$ is small, this matrix reduces to
\begin{eqnarray}
R_{\mathrm{rf}}\approx  
\left( 
\begin{array}{cc}
1 & 0 \\
-\frac{i\omega\tilde\gamma^{\prime}}{2cp_i}\Delta z & 1
\end{array} 
\right).
\end{eqnarray}
From equation Eq.~(\ref{LarmorEOM}), it is clear that this is nothing but a thin lens approximation to the focusing delivered by the RF magnetic field.  The complete matrix for the step $\Delta z$ is found by including the effects of the field edges:
\begin{eqnarray}
\lefteqn{\Delta M_{x,x^{\prime}}^{\mathrm{rf}} = R_E^{-1}\left(\gamma_f,\tilde\gamma^{\prime}e^{i\omega\Delta t}\right)R_{\mathrm{rf}}(\gamma_f,\tilde\gamma^{\prime},e^{i\omega\Delta t})
M_{i\rightarrow f}^{\mathrm{dc}}R_E(\gamma_i,\tilde\gamma^{\prime})}
\nonumber
\\
&=&
\left( 
\begin{array}{cc}
1 & 0 \\
\frac{\mathrm{Re}[\tilde\gamma^{\prime}e^{i\omega (t_i+\Delta t)}]}{2\gamma_f\beta_f^2} & 1
\end{array} 
\right)
\left( 
\begin{array}{cc}
1 & 0 \\
\frac{1}{2\gamma_f}\mathrm{Re}[\tilde\gamma^{\prime}(1-e^{i\omega\Delta t})e^{i\omega t_i}]& 1
\end{array} 
\right)
 \left( 
\begin{array}{cc}
C &\frac{p_i}{b}S\\
-\frac{b}{p_f}S & \frac{p_i}{p_f}C
\end{array} 
\right)
\left( 
\begin{array}{cc}
1 & 0 \\
-\frac{\mathrm{Re}[\tilde\gamma^{\prime}e^{i\omega t_i}]}{2\gamma_i\beta_i^2} & 1
\end{array} 
\right).
\label{transferMat}
\end{eqnarray} 
This is the main result of this work.  As a reminder, $\tilde\gamma^{\prime} = e\tilde\mathcal{E}_z(z_i)/mc^2$, where $\tilde\mathcal{E}_z$ is the complex electric field map. Additionally, the expressions for $\gamma$, $p$, and $\Delta t$ can be found in Eqs.~(\ref{gamma}-\ref{Leqn}), respectively.  For clarity we leave the result in the above factorized matrix form.  This allows several limiting cases to be easily evaluated.  First, in the limit that both the RF and solenoid fields vanish, $b,\tilde\gamma^{\prime}\rightarrow0$, and Eq.~(\ref{transferMat}) reduces to a drift matrix.  For vanishing RF fields, $\tilde\gamma^{\prime}\rightarrow 0$, and Eq.~(\ref{transferMat}) reduces to a hard edge solenoid matrix.  In the limit that $\omega\rightarrow 0$, $R_{\mathrm{rf}}$ reduces to $I_{2\times 2}$, and the total RF matrix reduces to the previous electrostatic result in Eq.~(\ref{DCmatrix}).  In addition to having the correct limiting behavior, the RF transfer matrix also has the correct determinant: $\mathrm{det}[\Delta M_{x,x^{\prime}}^{\mathrm{rf}}]= \mathrm{det}[\Delta M_{x,x^{\prime}}^{\mathrm{dc}}] = p_i/p_f$.  This follows directly from the fact that $\mathrm{det}[R_{\mathrm{rf}}] = 1$.

\section{Testing the Transfer Matrix}

To test the validity of our approach, the energy gain and transfer matrix are calculated through three different field set-ups and compared to direct integration of the equations of motion using a fourth order Runge-Kutta algorithm.  The three field set-ups used are: a DC gun and  with overlapping solenoid, a SRF cavity, and a RF gun with a solenoid immersed cathode.  To check that the transfer matrix correctly describes the transverse dynamics in each case, all four transfer matrix elements are compared with Rugne-Kutta integration.    To do so, the two principle trajectories through each field set-up are computed.  These trajectories are defined by the initial phase space coordinates
$\mathbf{u}_1 = \left( 
\begin{array}{cc}
1 \hspace{0.1cm}\mathrm{mm}, & 0 
\end{array} 
\right)^{\mathrm{T}}$, and $\mathbf{u}_2 = \left( 
\begin{array}{cc}
0,  & 1 \hspace{0.1cm} \mathrm{mrad} 
\end{array} 
\right)^{\mathrm{T}}$.
Table-\ref{tab:simparam} shows the various settings used in each simulation.
\begin{table}[htb]
\caption{Simulation Parameters}\label{tab:simparam}
\begin{ruledtabular}
\begin{tabular}{ c | cccccc}
Field Set-up & Voltage & Phase & Max. $B_{\mathrm{sol}}$ & KE$(z=0)$ & Step Size Type &Step Size  \\
\hline
DC gun \& Solenoid & 500 kV & 0 & 0.04 T & 1 eV & fixed & 0.1 mm \\
\hline
SRF Cavity & 3 MV & on-crest & N/A & 1 MeV & fixed & 2 mm 
\\
\hline
RF gun \& Solenoid & 1 MV & on-crest & 0.04 T & 1 eV & adaptable & 0.1 mm (avg.) \\
\end{tabular}
\end{ruledtabular}
\end{table}

\begin{figure}[ht!]
    \begin{center}
        \subfigure[\hspace{0.2cm}Field maps for the dc gun and solenoid.]{%
            \label{fig:dcpic}
            \includegraphics[width=0.5\textwidth]{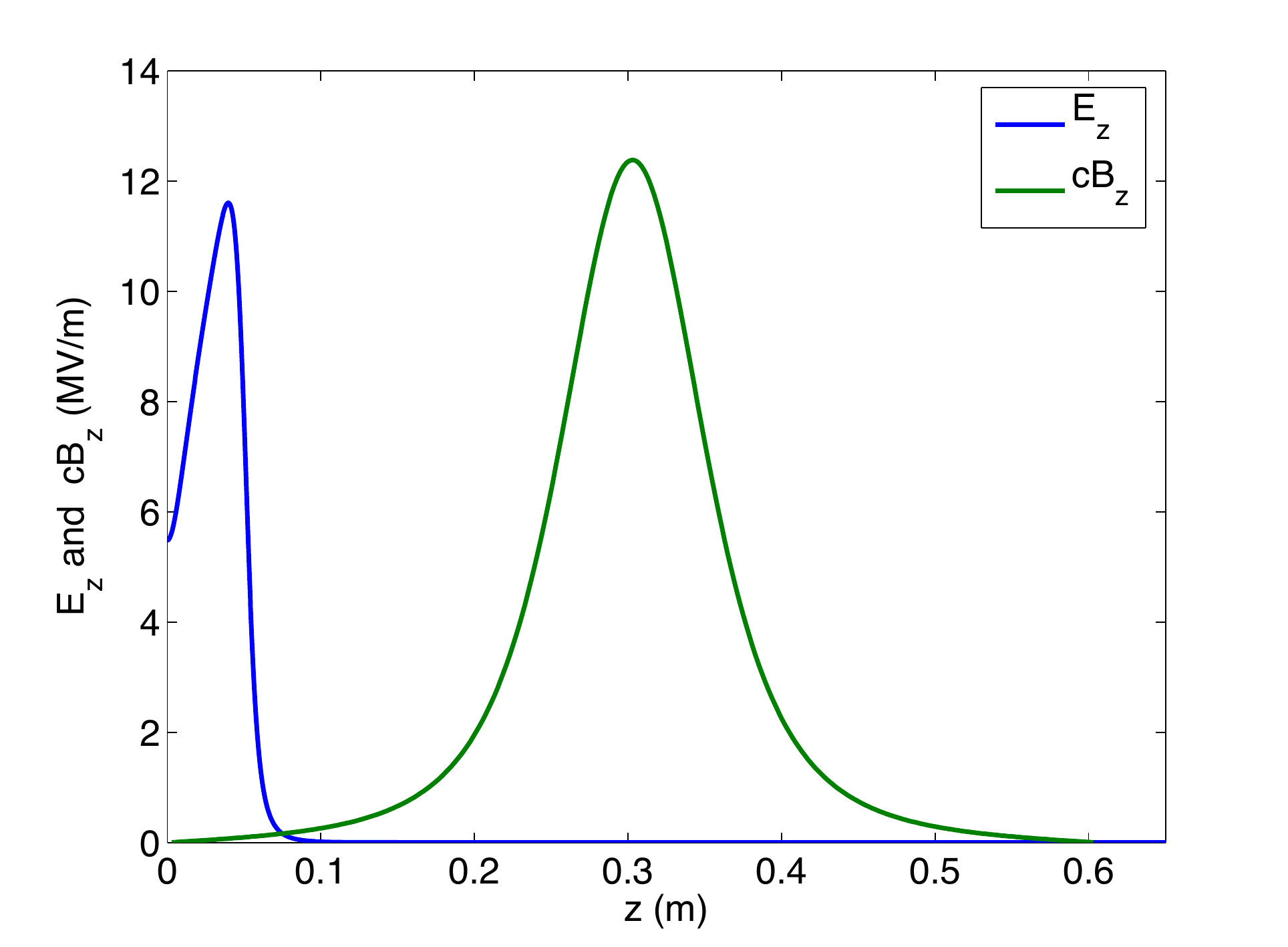}
        }%
        \subfigure[\hspace{0.2cm}Energy gain through the dc gun.]{%
           \label{fig:dcke}
           \includegraphics[width=0.5\textwidth]{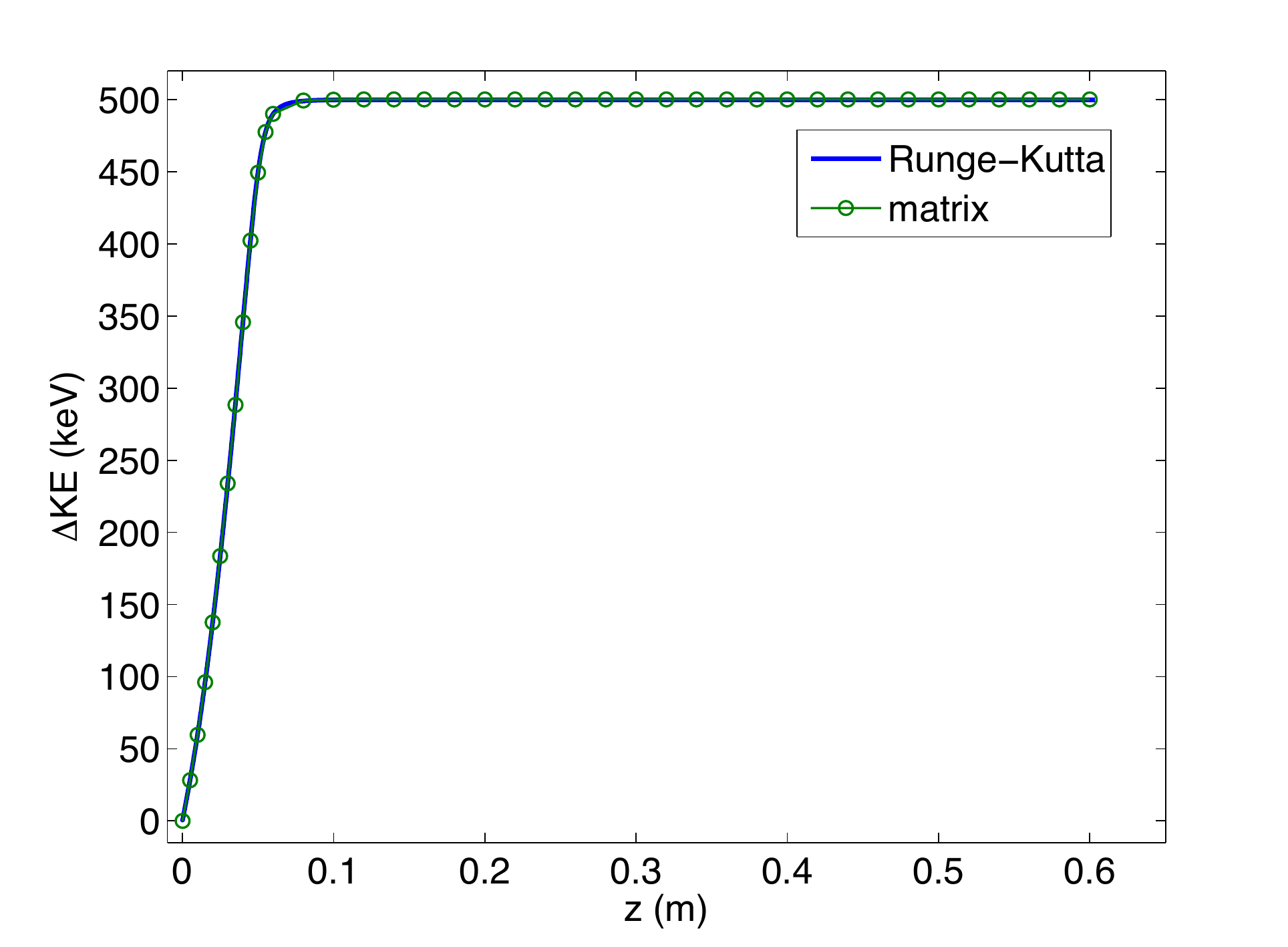}
        }
    \end{center}
    \caption{%
    \label{fig:DCpics}  The fields and energy gain for a 500 kV gun voltage, 0.04 T maximum solenoid field setting, and 1 eV initial kinetic energy.
     }%
\end{figure}

\begin{figure}[tb]
    \centering
    \includegraphics*[width=130mm]{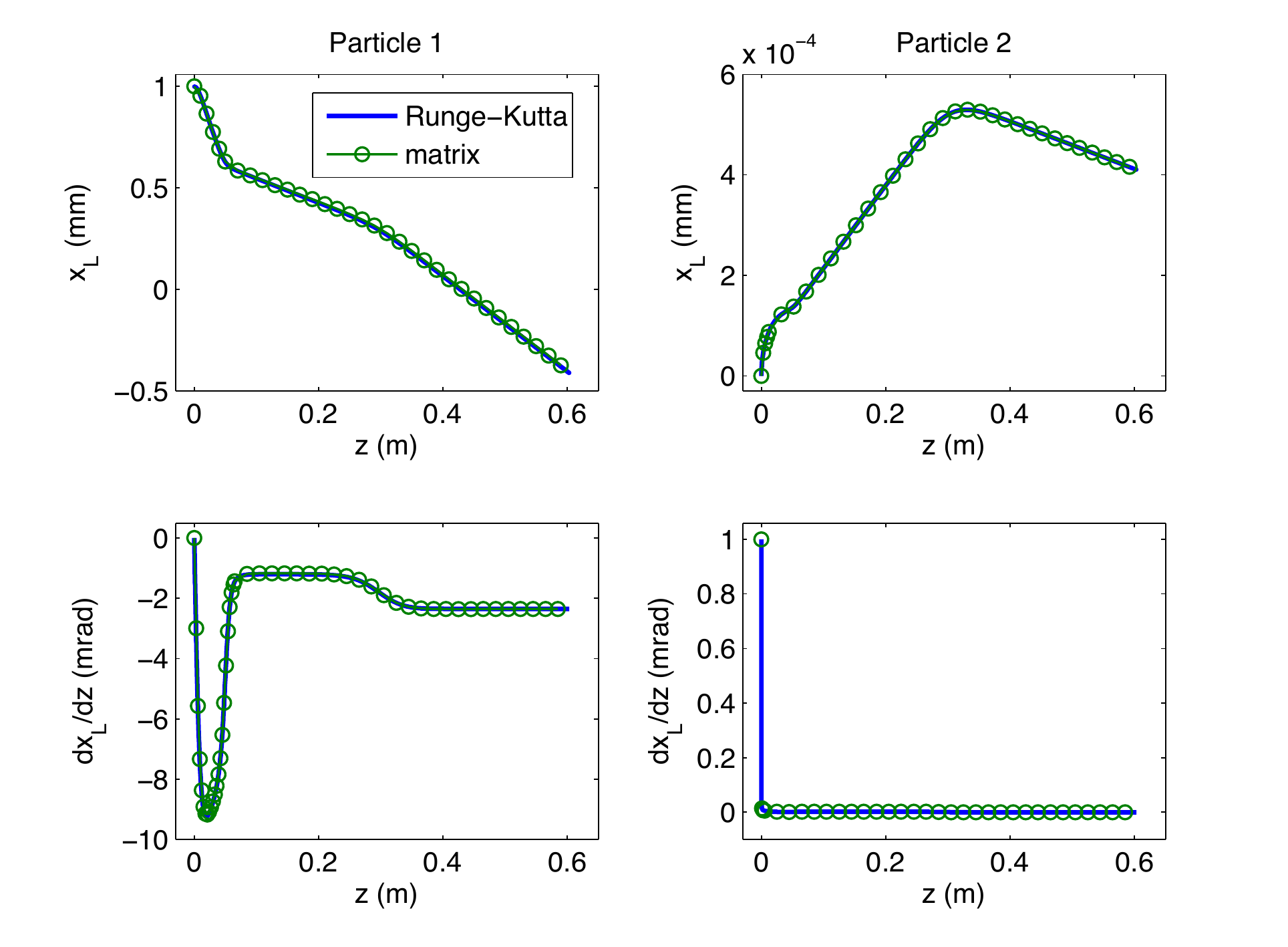}
    \caption{Comparison of Runge-Kutta integration (blue) and the tracking using the transfer matrix (green) through the DC gun and solenoid fields.}
    \label{fig:DCgun}
\end{figure}

\begin{figure}[ht!]
    \begin{center}
        \subfigure[\hspace{0.2cm}SRF cavity field map.]{%
            \label{fig:cavpic}
            \includegraphics[width=0.48\textwidth]{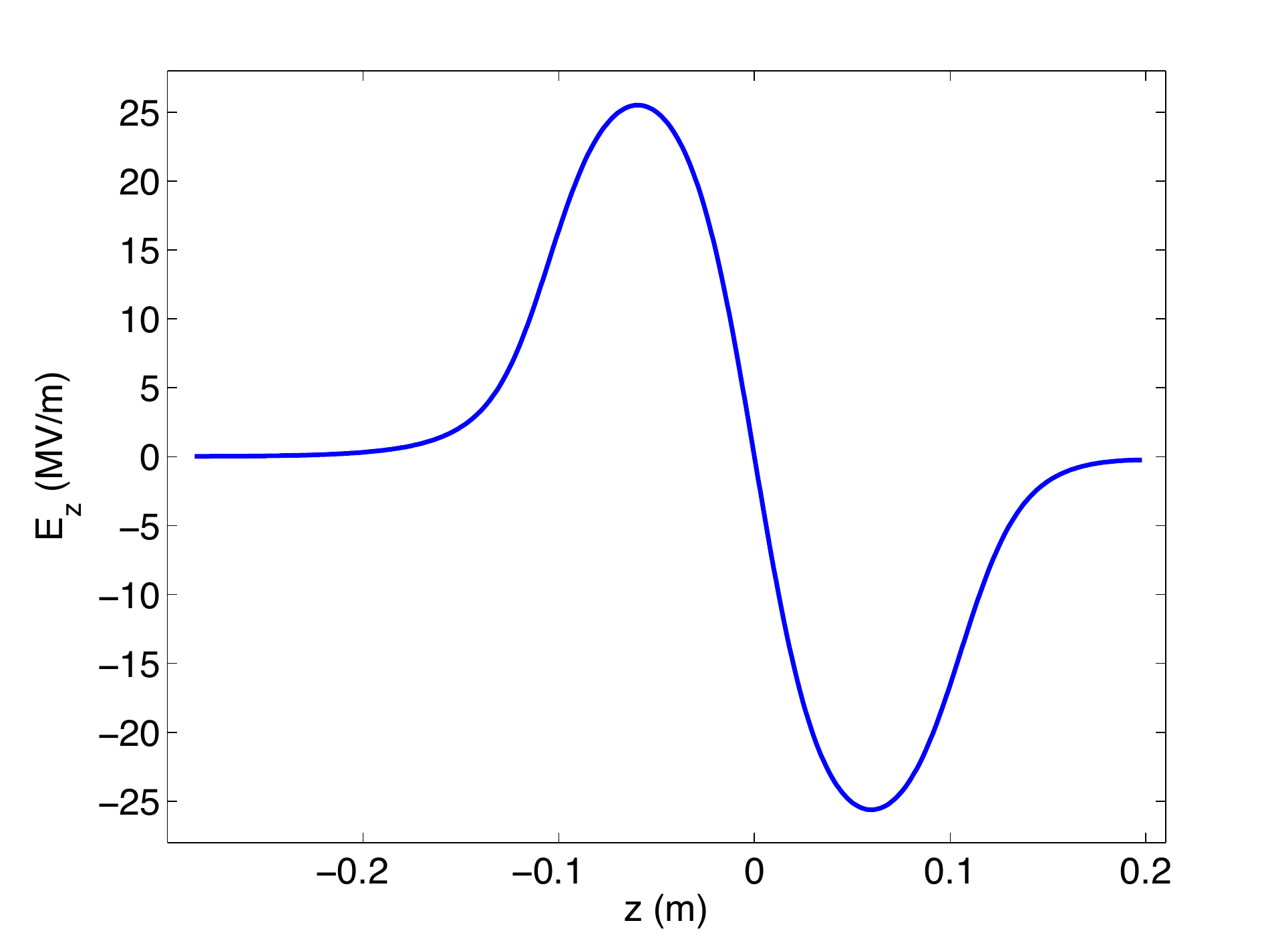}
        }%
        \subfigure[\hspace{0.2cm}Energy gain through the SRF cavity.]{%
           \label{fig:rfke}
           \includegraphics[width=0.48\textwidth]{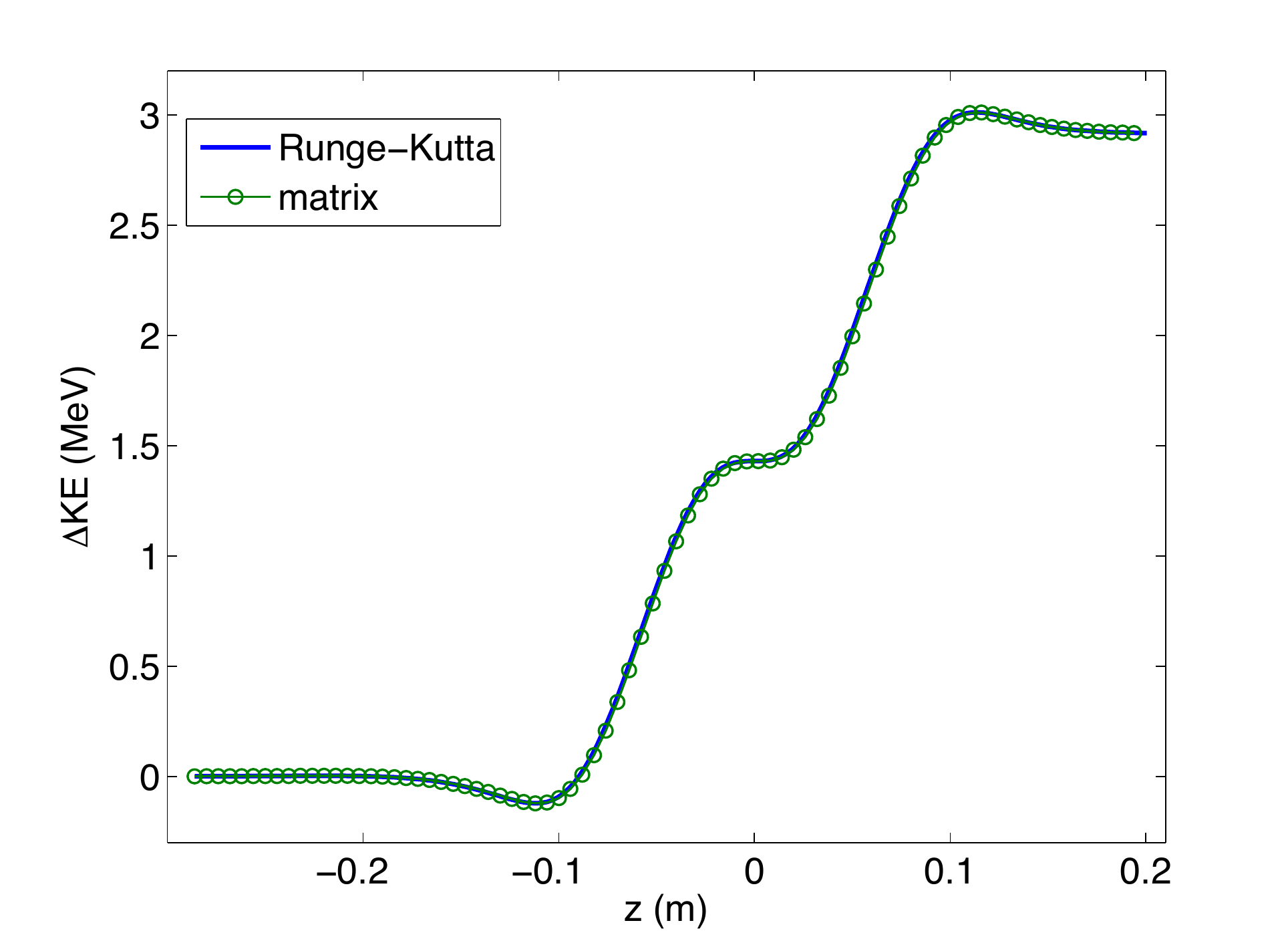}
        }
    \end{center}
    \caption{%
    \label{fig:RFpics}
        The field map for the SRF injector cavity and the corresponding energy gain.  The cavity voltage is 3 MV, the initial kinetic energy is 1 MeV, and the phase is on-crest.
     }%
\end{figure}

\begin{figure}[tb]
    \centering
    \includegraphics*[width=130mm]{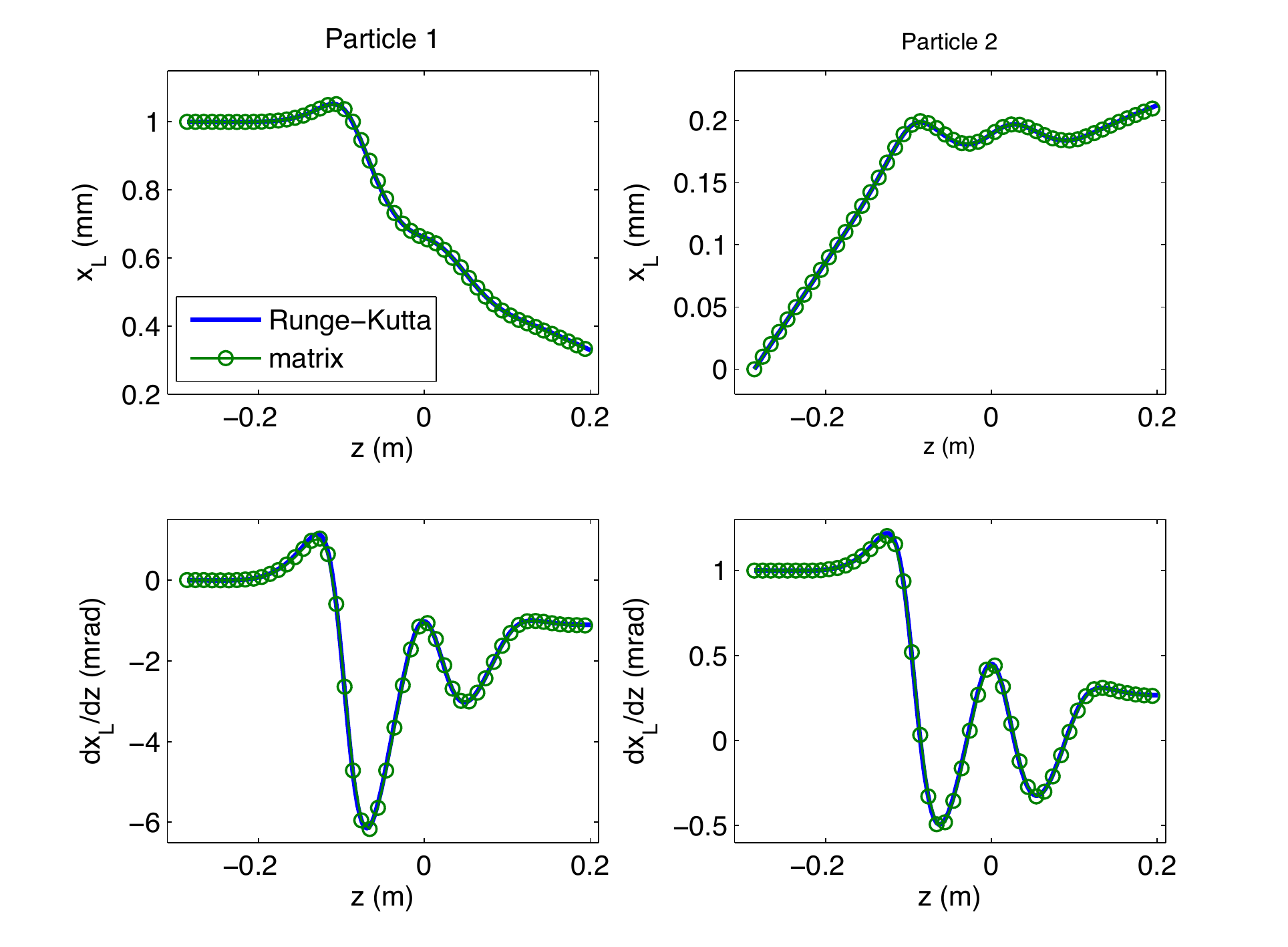}
    \caption{Comparison of direct integration (blue) and tracking using the transfer matrix (green) of the two principle trajectories.}
    \label{fig:rftrack}
\end{figure}

\begin{figure}[tb]
    \centering
    \includegraphics*[width=130mm]{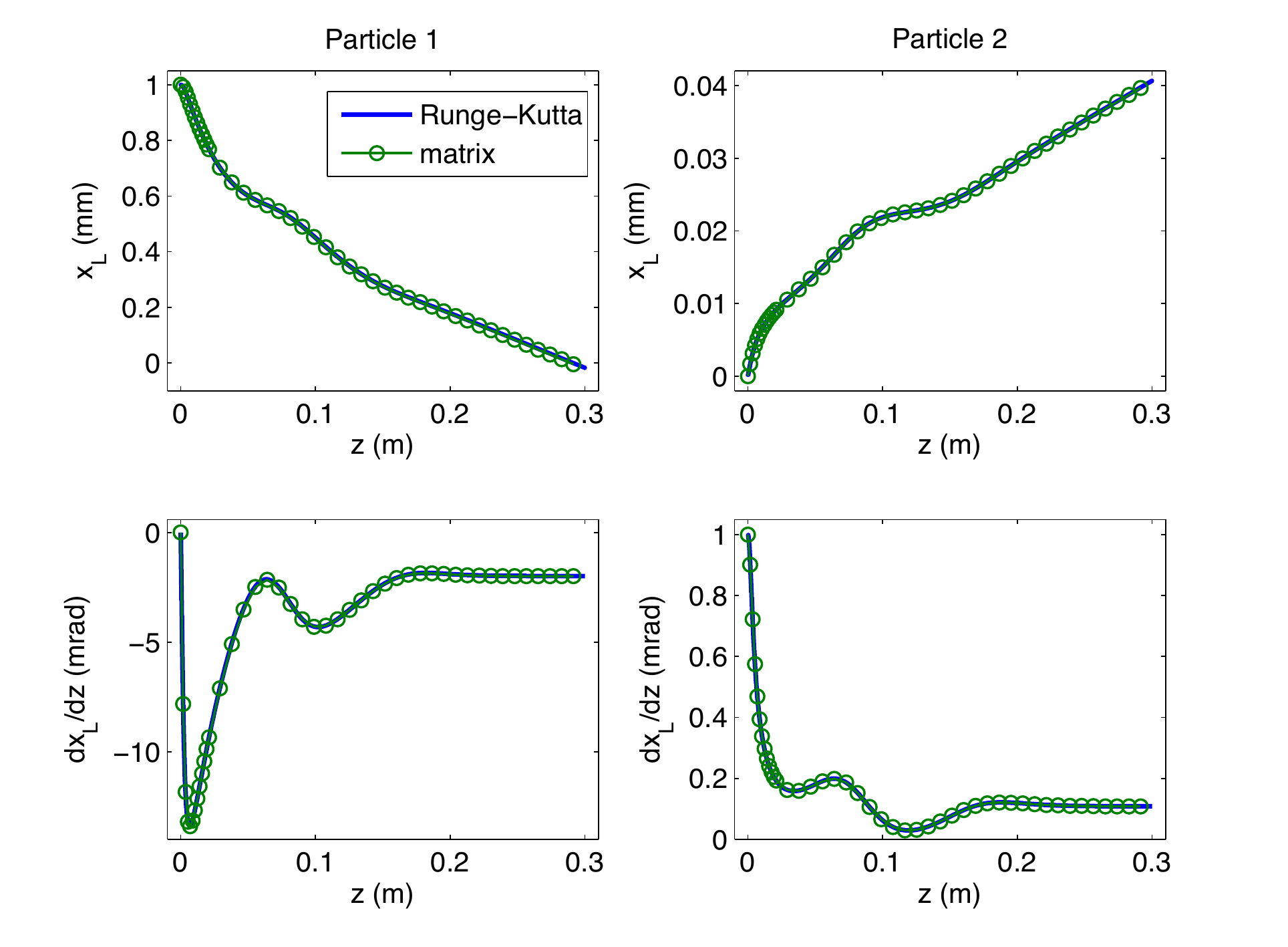}
    \caption{Comparison of direct integration (blue) and tracking using the transfer matrix (green) of the two principle trajectories.}
    \label{fig:RFgun}
\end{figure}

For the first simulation, we use the field maps for the the high voltage DC gun and first emittance compensation solenoid of the Cornell ERL injector prototype.  Fig.~\ref{fig:dcpic} shows the field maps corresponding to gun voltage and solenoid field strength given in Table-\ref{tab:simparam}.  Fig~\ref{fig:dcke} shows how the energy gain computed from the constant field solution in Eq.~(\ref{gamma}) compares to the energy gain computed using Runge-Kutta integration.  The step sized used for the constant field solution is 0.1 mm.  The agreement between the two methods demonstrates that the constant field solution works very well for the longitudinal variables, even for low initial kinetic energies.  
Fig.~\ref{fig:DCgun} shows the results of tracking the principle trajectories using both the RF transfer matrix with $\omega= 0$, and Runge-Kutta integration. 
As with with longitudinal variables, the agreement between both methods of tracking is excellent.  In addition to these results, the expression for the electrostatic transfer matrix has been experimentally verified in \cite{ivan1}.  Next, the two principle trajectories are computed through the field map of the 1.3 GHz Cornell ERL injector SRF cavity. 
Fig.~\ref{fig:cavpic} shows the on-axis electric field map of the SRF cavity model with a 3 MV cavity voltage.   
The corresponding energy gain through the cavity computed using the constant field solution and Runge-Kutta integration are shown in Fig.~\ref{fig:rfke}.  The results of tracking the two principle trajectories through the cavity with a fixed step size of 2 mm are shown in Fig~\ref{fig:rftrack}.
As in the electrostatic case, the agreement is very good.  Finally, the principle trajectories are computed through the RF gun set-up.  To simulate an RF gun, the last 1.5 cells of the injector cavity field map are used.   The solenoid field is positioned so that the maximum value of the solenoid field occurs at the cathode.  In order to make sure that the RF phase is constant over each step, a simple adaptive step size algorithm is included.  This algorithm adjusts the step size so that the change in RF phase over the step is less than a user defined tolerance.
\begin{figure}[ht!]
    \begin{center}
        \subfigure[\hspace{0.2cm}RF gun field maps.]{%
            \label{fig:rfpic}
            \includegraphics[width=0.48\textwidth]{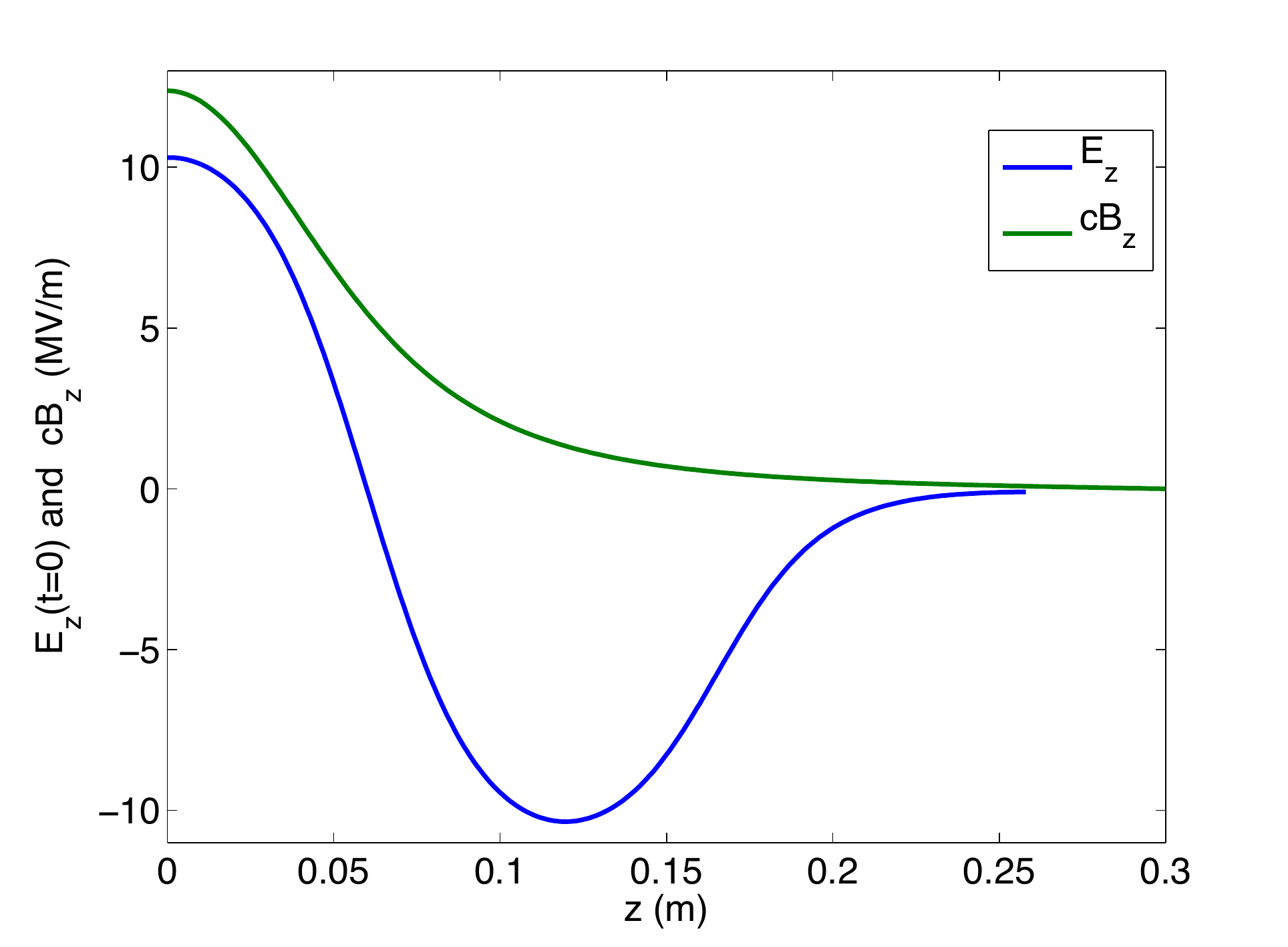}
        }%
        \subfigure[\hspace{0.2cm}Energy gain through the RF gun.]{%
           \label{fig:rfkegun}
           \includegraphics[width=0.48\textwidth]{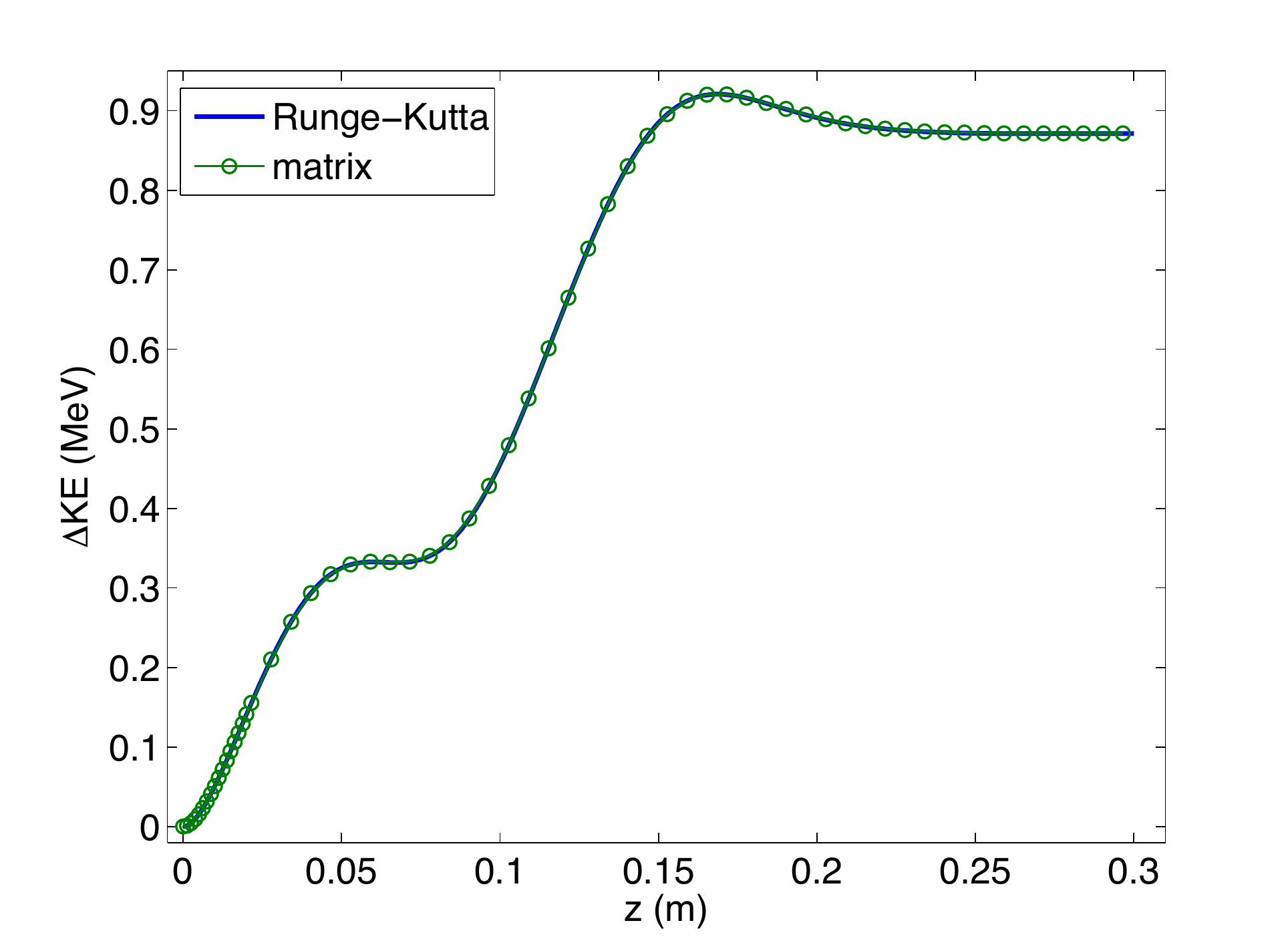}
        }
    \end{center}
    \caption{%
    \label{fig:RFpics}
       The fields and energy gain .  The cavity field is scaled and rotated so that the cavity voltage is 1 MV, and the phase is set to the on-crest value for a 1 MeV electron.
     }%
\end{figure}
Fig.~\ref{fig:rfpic} shows field maps for the RF gun set-up.  The corresponding energy gain through the gun is shown in Fig.~\ref{fig:rfkegun}.  The accelerating field is scaled so that the RF cavity voltage is 1 MV and the phase is set for maximum acceleration.  The tracking results for the two principle trajectories are shown in Fig.~\ref{fig:RFgun}.
From the figure, it is clear that that the transfer matrix works well in the low energy case.  The average step size for the simulation was roughly $\Delta z = 0.1$ mm.

\section{Conclusion}

We have derived and tested a new method for calculating the $4\times4$ transfer matrix through superimposed RF and solenoid fields.  The algorithm computes the transfer matrix directly from the field data without computing eigenfunction expansions or numerical derivatives.  Comparison with numerical integration demonstrates that this new method works for low energy beams starting from a solenoid immersed cathode.  Additionally, because the algorithm relies on an analytic solutions to the equations of motion, it is simple to implement and guarantees the correct value for the determinant of the transfer matrix.  One limitation to this approach is the assumption (inherent in the derivation) that the fields display cylindrical symmetry.  For many applications this is a reasonable assumption, however previous work shows that asymmetric focusing from input power couplers may be noticeable when heavy beam loading is present \cite{col}.  In addition, when tracking ultra-relativistic particles, the algorithm takes steps typically on the order of a few millimeters, and therefore may not be the best choice for computational speed.  In this case, one may still choose to use the Rosenzweig-Serafini matrx.  Nonetheless, the matrix algorithm given here strikes an appropriate balance between accuracy, speed, and simplicity not previously achieved.

\begin{acknowledgments}
This work is supported by NSF award DMR-0807731.
\end{acknowledgments}

\end{document}